**Hotspots and Trends in Magnetoencephalography Research (2013–2022): A Bibliometric Analysis**

Shen Liu[1,*], Jingwen Zhao[2]

[1]Department of Psychology, School of Humanities and Social Sciences, Anhui Agricultural University, No. 130 Changjiang Road(W), Shushan District, Hefei 230036, China

[2]Department and Institute of Psychology, Ningbo University, Ningbo, No. 616 Fenghua Road, Jiangbei District, Ningbo 315211, China

**Corresponding author**

Shen Liu

Department of Psychology, School of Humanities and Social Sciences, Anhui Agricultural University, Hefei, Anhui, 230036, China

E-mail: liushen@ahau.edu.cn

**Acknowledgement**

This work was supported by the Outstanding Youth Program of Philosophy and Social Sciences in Anhui Province (2022AH030089) and the Starting Fund for Scientific Research of High-Level Talents at Anhui Agricultural University (rc432206).

**Author Contributions**

S.L. conceived and designed the study. S.L. and J.Z. wrote and revised the main manuscript. S.L. and J.Z. collected and analyzed the data. S.L. and J.Z. prepared all figures and tables. All authors reviewed the manuscript.

**Declarations of interest**

None.

**Data availability**

The datasets generated and analyzed during the current study are not publicly available. The datasets are available from the corresponding author on reasonable request when the aim is to verify the published results.

**Hotspots and Trends in Magnetoencephalography Research (2013–2022): A Bibliometric Analysis**

**Abstract:** This study aimed to utilize bibliometric methods to analyze trends in international Magnetoencephalography (MEG) research from 2013 to 2022. Due to the limited volume of domestic literature on MEG, this analysis focuses solely on the global research landscape, providing insights from the past decade as a representative sample. This study utilized bibliometric methods to explore and analyze the progress, hotspots and developmental trends in international MEG research spanning from 1995 to 2022. The results indicated a dynamic and steady growth trend in the overall number of publications in MEG. Ryusuke Kakigi emerged as the most prolific author, while *Neuroimage* led as the most prolific journal. Current hotspots in MEG research encompass resting state, networks, functional connectivity, phase dynamics, oscillation, and more. Future trends in MEG research are poised to advance across three key aspects: disease treatment and practical applications, experimental foundations and technical advancements, and fundamental and advanced human cognition. In the future, there should be a focus on enhancing cross-integration and utilization of MEG with other instruments to diversify research methodologies in this field.

## 1. Introduction

Magnetoencephalography (MEG) is a non-invasive technique used to detect electromagnetic physiological signals in the brain. It offers millisecond-level time resolution and millimeter-level spatial resolution, unaffected by factors such as tissue conductivity and skull thickness. MEG can study neural activity across various frequency bands, offering clearer insight into brain function (Baillet, 2017).

Compared to neuroimaging techniques like functional magnetic resonance imaging (fMRI) and positron emission computed tomography (PET), which are based on hemodynamics, MEG offers a more direct and concise method for measuring brain activity. It is widely utilized for functional localization and assessment of various brain regions (Kharkar & Knowlton, 2015). Additionally, MEG offers a superior method for measuring brain activity and is extensively employed in the functional localization and assessment of various brain regions (Kharkar & Knowlton, 2015). The primary challenge limiting the applicability of MEG is the need to achieve sufficient sensitivity for detecting the small magnetic fields (approximately 100 fT) generated by the brain, necessitating the use of sensor coils coupled to SQUID in current MEG systems (Cohen, 1972). Since the sensors typically need to be cooled to about –269°C, several challenges arise: first, the sensors must form a fixed array around the scalp; second, MEG signal strength decreases with the square of the distance from the source; third, the helmet design is optimized for individuals with relatively large heads, making it difficult to achieve uniform coverage and high sensitivity simultaneously for smaller heads. Lastly, the system's complexity contributes to its high acquisition and maintenance costs (Brookes et al., 2022). In recent years, a new magnetic field sensing technology has emerged in the field of MEG. OPMs are magnetic field sensors with sensitivity comparable to SQUID but do not rely on cryogenic cooling. This has spurred the development of new MEG systems; OPMs, which do not require cryogenic cooling like SQUIDs, are enabling "OPM-MEG" scanners. These systems are still in early technological stages but promise advancements such as higher quality data, improved coverage uniformity, motion robustness, and reduced system complexity (Johnson et al., 2013).

Overall, MEG is primarily employed in cognitive neuroscience, playing a crucial role in elucidating the functions and mechanisms of the mind and brain. Understanding the brain's functional connectivity mechanisms represents a core research focus within this field. In recent years, various neuroimaging techniques including event-related potentials (ERPs), electroencephalography (EEG), fMRI, functional near-infrared spectroscopy (fNIRS), MEG, and PET have become indispensable tools in cognitive neuroscience research. With these techniques, substantial amount of data reflecting brain activity can be generated, further enhancing our understanding of complex cognitive functions (Hauk, 2020). Due to its millisecond-level time resolution and millimeter-level spatial resolution—superior to EEG for spatial resolution and to fMRI and fNIRS for time resolution—MEG is increasingly becoming a mainstream research tool in cognitive neuroscience.

Currently, MEG is widely utilized by researchers across various fields such as medicine, psychology, and neuroscience, yielding numerous significant research findings. However, the dispersed nature of research makes it challenging to objectively and comprehensively identify and explore the hotspots and developmental trends in MEG research. Bibliometrics, as a quantitative analysis method, offers a range of objective indicators to visualize changes in trends and relationship networks such as journals, institutions, authors, countries, references, keywords, etc. (Yan et al., 2020). Bibliometrics can assist researchers in exploring research hotspots and future development trends within a specific field, making it a primary tool for knowledge discovery in studying developmental trends and changes within fields (Pan et al., 2021). Some researchers have employed bibliometric methods to study EEG, fMRI, fNIRS, PET, and other technologies; however, there has been no research analyzing the research hotspots and development trends of MEG using bibliometric methods (Yan et al., 2020). Therefore, this study aims to utilize bibliometric methods to visualize international MEG research trends. Due to the limited volume of domestic literature on MEG, the analysis focuses solely on the global research landscape, using data from the past decade as an illustrative example.

## 2. Methods

### 2.1 Data acquisition and search strategy

Compared to other databases such as PubMed and Scopus, Web of Science (WOS) is regarded as the most authoritative, influential, and academically valuable database within the international academic community. Its retrieval system enjoys high reputation both domestically and internationally, thus we chose WOS as the database for our research. To ensure the accuracy and credibility of the literature, data analysis was conducted using the citation indexes of SSCI and SCI in the WOS Collection as of September 20, 2023. Advanced search was conducted using the search term "Magnetoencephalography" in the topic field. The publication timeframe was set from January 1, 1995 to December 31, 2022, with document types limited to "Article" and "Review Article", and language restricted to "English". After excluding irrelevant literature in this field, a total of 8,331 articles were retrieved.

### 2.2 Analysis tool

The retrieved dataset of 7,523 literature entries underwent visual analysis using CiteSpace (6.1.R6), VOSviewer (1.6.18), and Scimago Graphica (1.0.26). CiteSpace integrates and analyzes information on journals, authors, countries, institutions, keywords, and other parameters such as publication counts, citation metrics, impact factors, and average publication year (Chen, 2006). VOSviewer was utilized to construct maps of authors, journals, and keywords based on co-citation data (Van Eck & Waltman, 2010). Scimago Graphica was employed to generate geographic visualization views, offering insights into international collaborations (Kovačić & Petrak, 2022).

## 3. Results

### 3.1 Annual publications, annual citations, and publication trends

The number of publications can indicate the trend of scientific research development, while citation counts reflect publication quality. Figure 1 illustrated the annual publication and citation counts of MEG research.

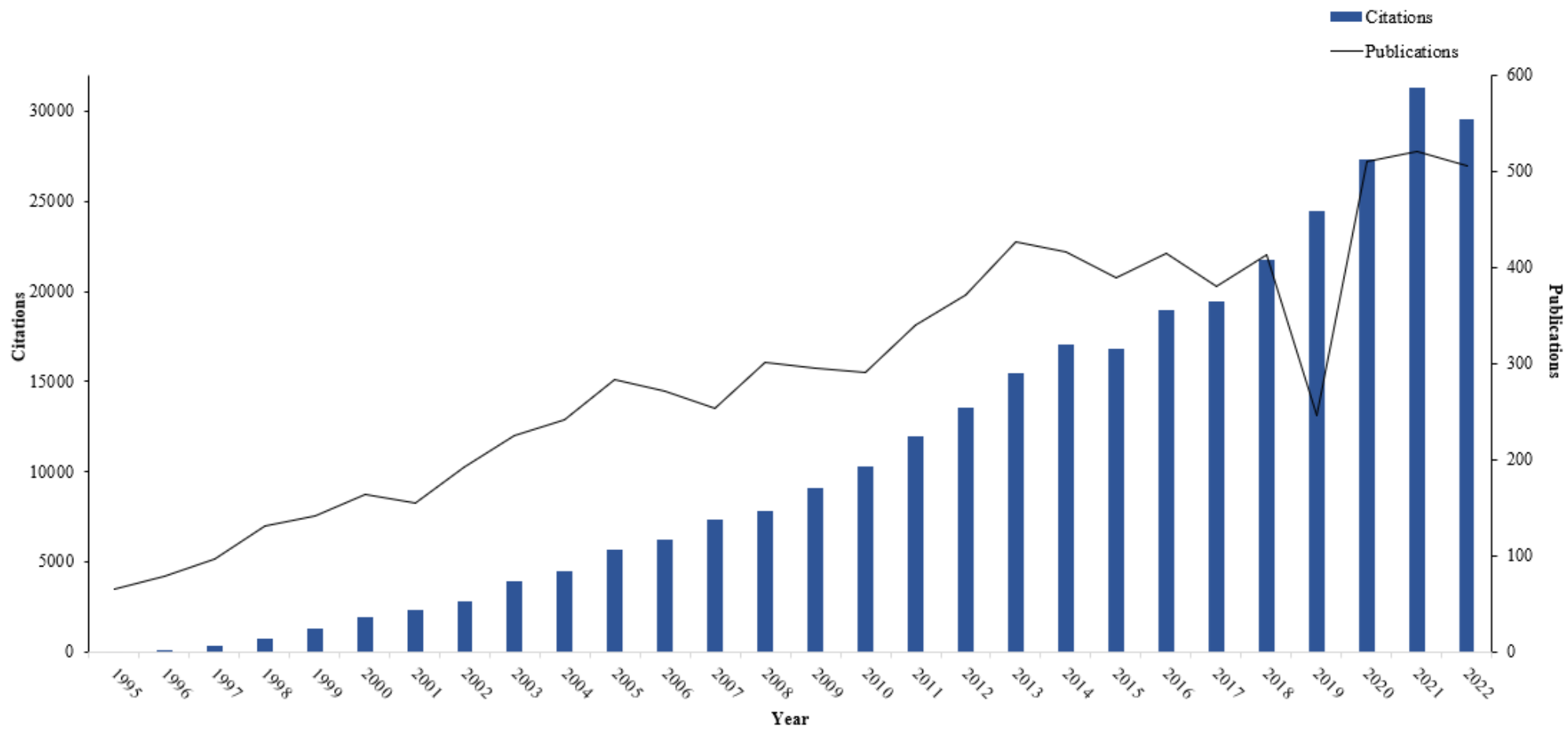

**Figure 1. Annual publications, annual citations, and publication trends of MEG research**

The number of publications increased from 65 in 1995 to 505 in 2022, demonstrating a dynamic and steady upward trend. This indicated sustained enthusiasm and dedication within the academic community towards MEG research (Ding & Yang, 2022). A total of 330,269 citations were accumulated for the 8,331literature entries, averaging 39.64 citations per paper. The most cited publication is Bullmore and Sporns' paper on graph theoretical analysis of structural and functional brain networks, published in *Nature Reviews Neuroscience* (Bullmore & Sporns, 2009), which has accumulated 7,150 citations to date. Among them, from 2018 to 2019, the number of published MEG documents exhibited a declining trend. This is attributable to the emergence of a new device called wearable MEG (Boto et al., 2018), which represents an innovation in MEG technology. Due to the innovative nature of the wearable MEG device, most researchers required a period of time to re-study the MEG equipment. Consequently, the number of published MEG documents decreased during this period. As researchers have completed a substantial number of studies on wearable MEG devices, the number of published MEG papers has increased rapidly from 2019 to 2020. The citation counts exhibited a linear increasing trend from 1995 to 2022, indicating that the quality of MEG research publications is good, and the number of citations per year is consistently growing. These findings also imply that scholars′ interest in MEG research is continuously increasing, and the future citation counts will continue to rise, suggesting that MEG research will persist in receiving attention from the academic community.

**3.2 Authors analysis**

Through the search, a total of 8,331 articles were found, involving 18,592 contributing authors. Due to the large number of author nodes, the decision was made to focus on the top 100 authors based on their number of publications. A clustering network of these top 100 authors was created using VOSviewer to identify potential collaboration relationships, as shown in Supplementary Figure 1. In Supplementary Figure 1, the clustering network consists of a total of 9 clusters and 100 nodes. The size of each node in the clustering network represents the number of publications by the respective author, while the connections between nodes depict the collaboration relationships among the authors. Detailed data regarding the top 10 authors based on publication count can be found in Table 1. The number of publications per author in the top 10 range from 98 to 195, with an average of 114.2 publications. Among the top 10 authors, Ryusuke Kakigi from the National Institute for Physiological Sciences has the highest publication count at 195 (2.34%), followed by Tony W. Wilson from the Boys Town National Research Hospital with 155 publications (1.86%), and Riitta K. Hari from Aalto University with 132 publications (1.58%). The H-index is an important metric used to measure the academic influence of authors. In this study, the H-index values for the authors were obtained by searching their paper indices on the WOS platform. Ryusuke Kakigi from National Institute for Physiological Sciences has the highest H-index at 101, followed by Alfons Schnitzler form Ruhr University Bochum with an H-index of 72, and Christo Pantev from University of Münster with an H-index of 65. These top three authors demonstrate a strong academic influence in the field of MEG research.

**Table 1. Top 10 authors on MEG research.**

| Rank | Author | Institution | Publication (%) | H–index |
|---|---|---|---|---|
| 1 | Ryusuke Kakigi | National Institute for Physiological Sciences | 195 (2.34) | 59 |
| 2 | Tony W. Wilson | Boys Town National Research Hospital | 155 (1.86) | 37 |
| 3 | Riitta K. Hari | Aalto University | 132 (1.58) | 101 |
| 4 | Fernando Maestu | Complutense University of Madrid | 132 (1.58) | 42 |
| 5 | Arjan Hillebrand | Amsterdam UMC Locat Vrije Univ Amsterdam | 123 (1.48) | 50 |
| 6 | Andrew C. Papanicolaou | University of Tennessee Health Science Center | 113 (1.36) | 49 |
| 7 | Timothy P. L. Roberts | Childrens Hospital of Philadelphia (United States of America) | 108 (1.30%) | 43 |
| 8 | Alfons Schnitzler | Ruhr University Bochum | 106 (1.27%) | 72 |
| 9 | Christo Pantev | University of Munster | 103 (1.24%) | 65 |
| 10 | Matthew J. Brookes | University of Nottingham | 98 (1.18) | 47 |

### 3.3 Journals analysis

After the search, a total of 8,331 articles were found, published across 839 different journals. Among the top 10 journals in terms of the number of published articles, the publication counts range from 151 to 950, with an average of 189.2 articles per journal. These top 10 journals account for 34.71% of the total number of articles published. As shown in Table 2, the journal with the highest number of published articles was *NeuroImage* with 950 articles, and it also had a relatively high impact factor of 7.400. This was followed by *Clinical Neurophysiology* with 325 articles and an impact factor of 4.700, and *Human Brain Mapping* with 309 articles and an impact factor of 4.800. Additionally, the impact factors of these top 10 journals ranged from 1.703 to 7.400, with an average of 4.23. While two of the top 10 journals have an impact factor over 5.000, the majority have impact factors below this threshold, indicating that publishing MEG research papers in very high-impact factor journals remains challenging. Furthermore, by utilizing the VOSviewer tool, a clustering visualization of the MEG research publishing journals was created, as depicted in Supplementary Figure 2. Due to the large number of journal nodes, the decision was made to focus the clustering visualization in Supplementary Figure 2 on the top 100 journals based on their number of published articles. The size of the nodes represents the number of articles published in the journal, and the color of the nodes indicates various fields of MEG research. The clustering visualization in Supplementary Figure 2 showed a total of 6 distinct clusters comprising 100 journal nodes. As indicated by the largest node, the journal *NeuroImage* had published the highest number of articles. The relationship networks between and within each of the 6 clusters appeared to be closely connected.

**Table 2 Top 10 journals in MEG research**

| Rank | Journals | Publications (%) | H–index |
|---|---|---|---|
| 1 | Neuroimage | 950 (11.40) | 7.400 |
| 2 | Clinical Neurophysiology | 325 (3.90) | 4.700 |
| 3 | Human Brain Mapping | 309 (3.71) | 4.800 |
| 4 | NeuroReport | 253 (3.04) | 1.703 |
| 5 | Cerebral Cortex | 196 (2.35) | 3.700 |
| 6 | Journal of Neuroscience | 194 (2.33) | 6.709 |
| 7 | PLoS ONE | 188 (2.56) | 3.752 |
| 8 | Neuroscience Letters | 168 (2.02) | 3.197 |
| 9 | frontiers in Human Neuroscience | 158 (1.90) | 2.900 |
| 10 | Brain Topography | 151 (1.81) | 3.400 |

### 3.4 Countries and institutions analysis

After the search, a total of 8,331 articles were found to have been published across 92 countries. Using the Scimago Graphica tool, a geographical visualization was created to depict the distribution of these countries, as shown in Figure 2. In the geographical visualization shown in Figure 2, the size of each node represents the number of articles published by that country. The interconnections between the nodes indicate relationships between the countries. Additionally, the node colors reflect different clusters of countries. The geographical visualization in Figure 2 clearly showed an uneven distribution of MEG research publications across countries. However, the United States stands out as having published far more articles than any other country, indicating that the US occupies a leading position in the field of MEG research. The geographical visualization in Figure 2 showed that Poland and Japan, both belonging to the purple cluster, had a broad cooperative network. Supplementary Table 1 provided specific details on the top 10 countries in terms of the number of published MEG research papers. The United States had the largest number of published papers at 2,796, accounting for 33.56% of the total. Germany was second with 1,573 papers (18.88%), followed by Japan with 1,300 papers (15.60%).

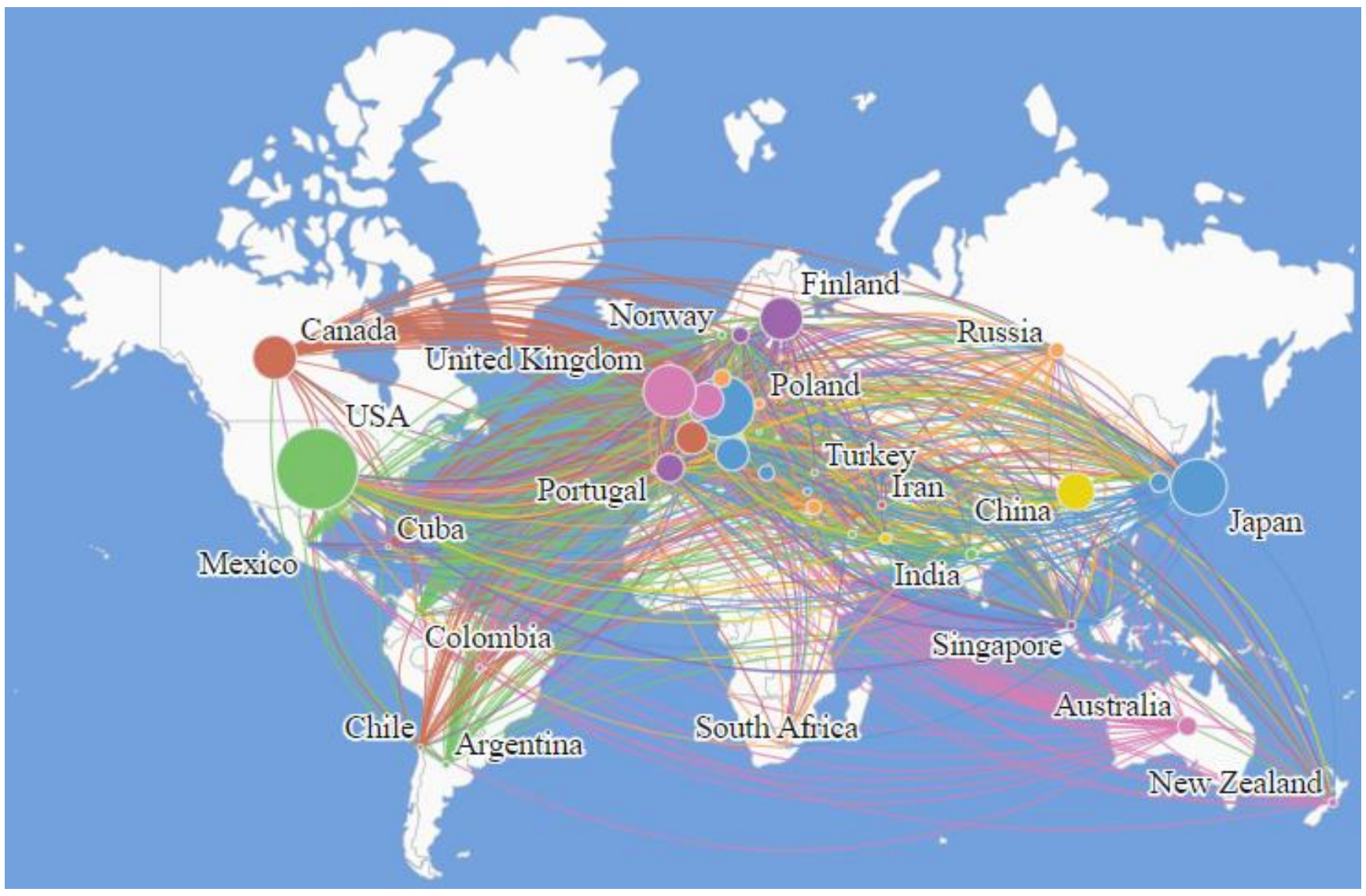

**Figure 2. Geographic visualization view**

After the search, a total of 8,331 articles were found to have been published across 4,072 different research institutions. Due to the large number of institutions nodes, the decision was made to focus the analysis on the top 100 institutions based on their number of published articles. Using the VOSviewer tool, a clustering visualization of these top 100 research institutions was created, as shown in Supplementary Figure 3. This clustering view comprises a total of 8 distinct clusters, each represented by 100 prominently featured nodes. As presented in Supplementary 2, the top 10 research institutions that have published the most MEG research papers were shown. The institutions with the highest number of publications were Aalto University with 538 papers (6.46%), followed by the University of Toronto with 465 papers (5.58%), and the University of Helsinki with 418 papers (5.02%).

### 3.5 Keywords analysis

Supplementary Figure 4 presented an analysis of the emerging keywords involved in MEG studies. The top 25 emerging keywords highlighted in this figure referred to terms that had shown a sudden increase in frequency over time, reflecting the frontier and development trends in this research field. As shown in Supplementary Figure 4, the red portion of each bar represented the time period when the keyword first emerged. The most frequently emerging keywords in the past four years included "connectivity" and "oscillation". Table 3 presented the top 10 keywords with the highest occurrence frequency in MEG research. The top three keywords were "eeg", "localization", and "response", indicating that MEG research tends to be used in conjunction with EEG, and is closely linked to the study of the brain. This finding indirectly illustrates the primary function and application of MEG research. Given that the brain is the central nervous system of humans, the high occurrence frequency of keywords like "brain" further demonstrates that MEG is fundamentally the study of the brain's function and activity.

**Table 3. Top 10 keywords in the frequency of magnetoencephalogram research**

| Rank | Keywords | Counts |
|---|---|---|
| 1 | eeg | 1094 |
| 2 | localization | 774 |
| 3 | response | 576 |
| 4 | activation | 505 |
| 5 | functional connectivity | 499 |
| 6 | perception | 444 |
| 7 | oscillation | 428 |
| 8 | potential | 382 |
| 9 | synchronization | 374 |
| 10 | fmri | 359 |

**3.6 Highly cited references analysis**

As shown in Supplementary Table 3, the top 10 most cited references in MEG research literature were presented. Among these, the paper by Bullmore Edward et al. (2009) had the highest citation counts on the WOS platform. This indicated that the cited references within this paper held high reference value, and its research achievements were significant and well-recognized by the broader scientific community. As illustrated in Supplementary Figure 5, the top 25 MEG-related papers with the strongest citation bursts over the past decade were presented. Among these, the paper by Sylvain Baillet (2017) had exhibited the strongest citation burst from 2017 to 2022, indicating that this paper held particularly high reference value in the cutting-edge research within the field of MEG.

**4. Hotspot analysis and trends prediction in MEG**

The 8,331 articles retrieved contained a total of 18,373 keywords, which reflected the core content and focus areas of the MEG research literature. Due to the large number of keyword nodes, the analysis was narrowed to focus on the top 100 most frequently occurring keywords. Using VOSviewer, a keyword clustering visualization was generated, as shown in Figure 3. This clustering view comprised 4 distinct clusters, each containing 100 nodes with distinctive features. In the keyword clustering visualization shown in Figure 8, the size of each node represented the frequency of that corresponding keyword. Larger nodes indicated keywords that were more commonly used in MEG research. The keyword "magnetoencephalography" appeared most frequently among the top 100 keywords, indicating its central and influential role within the MEG research field. In the keyword clustering visualization, the distance between any two nodes represents the relative strength and similarity between those keywords. Nodes of the same color are grouped into similar topical clusters. Due to space limitations, only the larger keyword clusters are highlighted in the study presented in this work. For instance, the red cluster contained keywords such as "synchronization", "working-memory", "attention" and "dynamics", suggesting this cluster is focused on the physiological mechanisms underlying human cognition and behavior. The green cluster contained keywords such as "magnetoencephalography", "stimulation", "cortex" and "human-brain", indicating this cluster is focused on research relating to brain functions and the application of MEG. The yellow cluster included keywords such as "mri", "eeg", "localization", and "source localizatiton", suggesting this cluster is focused on the applications of MEG and EEG, particularly in source localization. The blue cluster included keywords such as "perception", "responses", and "language", indicating this cluster was focused on different directions of cognitive processes.

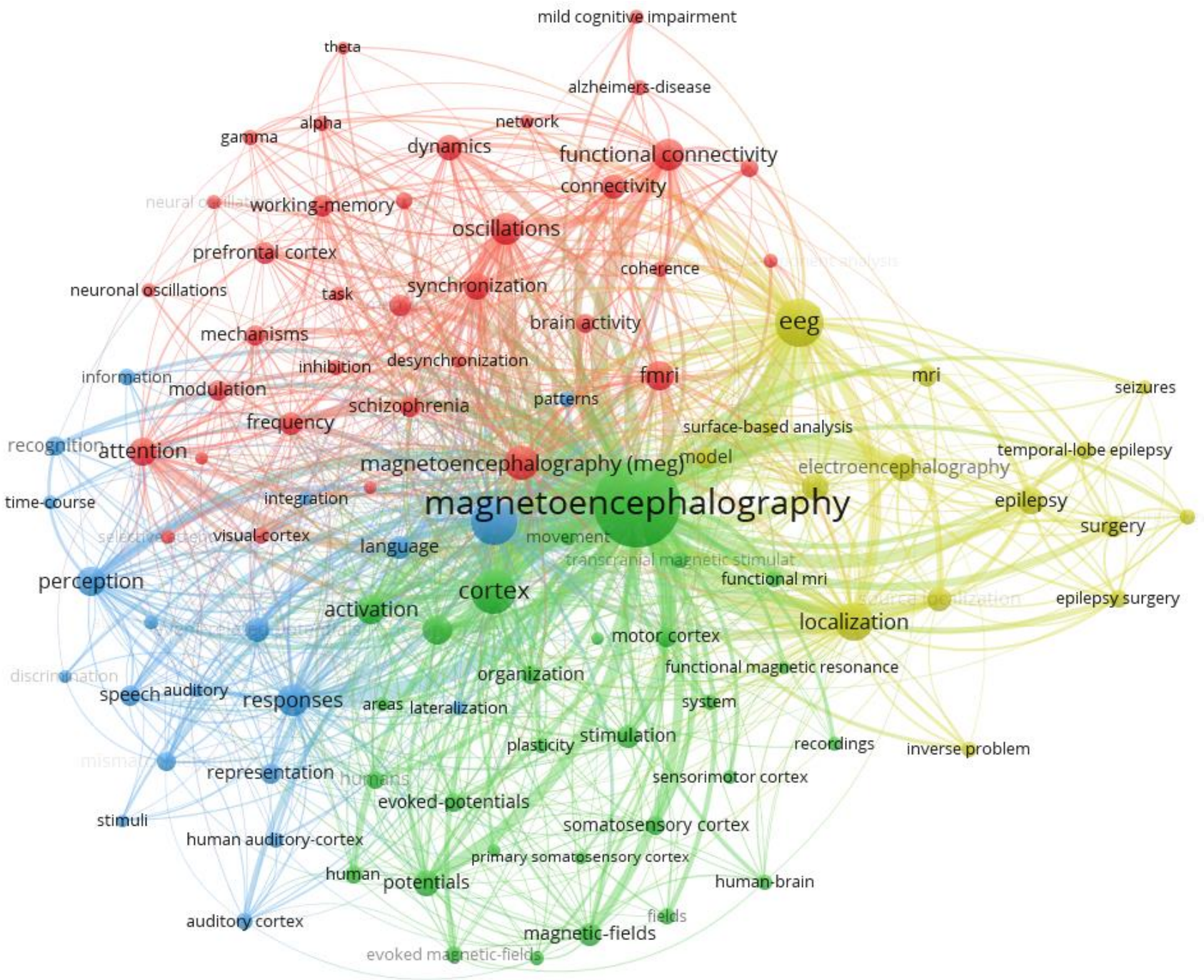

**Figure 3. Keywords in MEG study cluster view via VOSviewer**

Using CiteSpace, a deep clustering analysis was conducted on 2,563 keywords, resulting in the identification of 7 distinct clusters, as illustrated in Supplementary Figure 6. By adding a timeline dimension, the keyword cluster timeline view was generated, as shown in Figure 4. In the keyword cluster timeline view shown in Figure 4, the horizontal axis represented the publication year, while the vertical axis represented the different keyword clusters. In the keyword cluster timeline view, keywords from the same cluster are arranged horizontally in chronological order. Each cluster's keywords are displayed along the timeline axis corresponding to the time span of that particular cluster. By analyzing the timeline view of keyword clusters, researchers can explore the publication year, duration, and the rise, peak, and decline of specific clusters in research to investigate the temporal characteristics of different research areas (Chen, 2006). According to Chen (2006), when the cluster module value is greater than 0.3 and the average silhouette value is greater than 0.7, a cluster is considered significant and highly reliable when its module value exceeds 0.3 and its average silhouette value surpasses 0.7. In this study, the cluster module value was 0.4193, indicating a significant cluster, and the average silhouette value was 0.742, indicating a highly reliable cluster, thereby strengthening the credibility of the results.

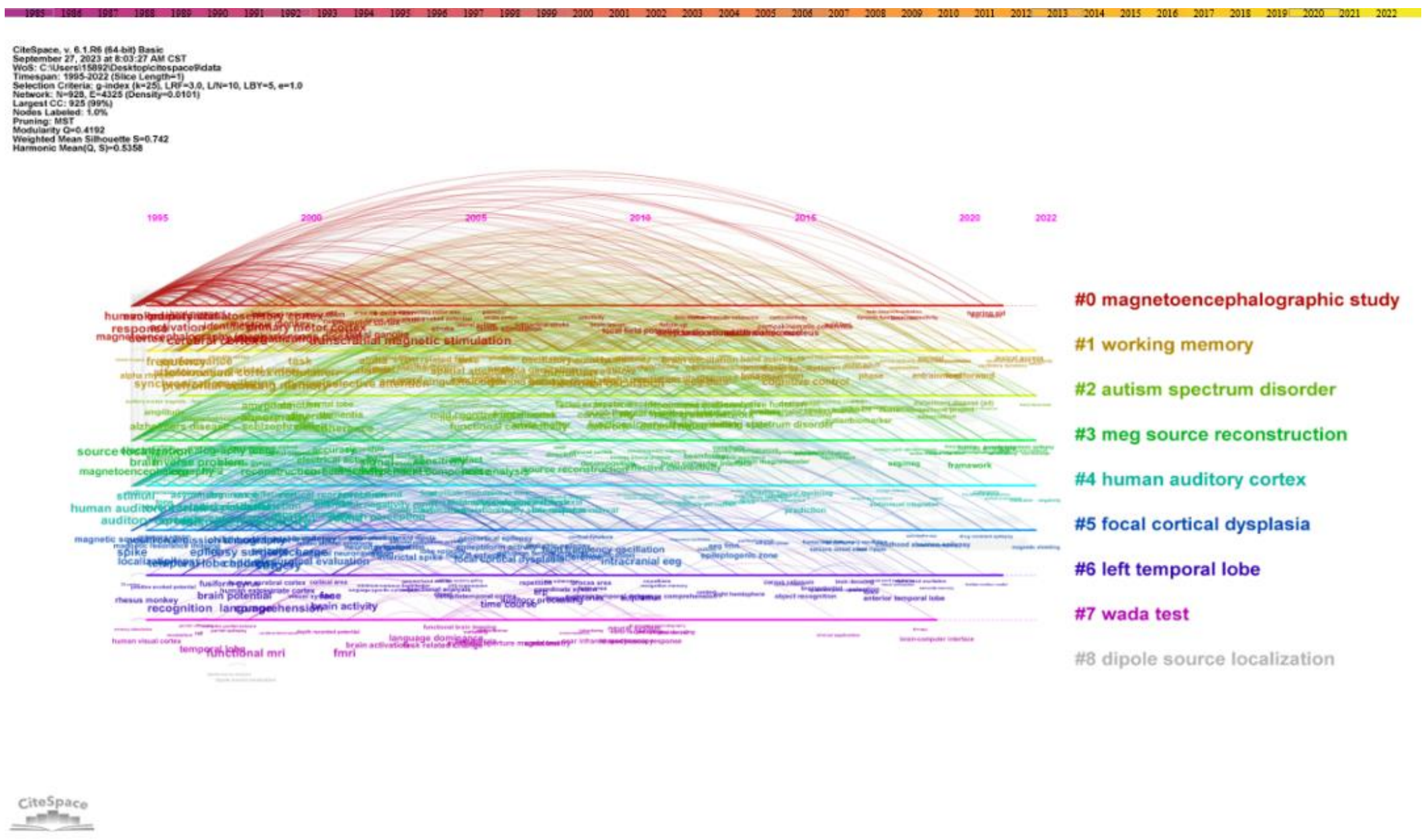

**Figure 4. Timeline view of co-occurrence of keywords**

Based on the findings in Table 4 and Figure 10, this study will analyze future trends and directions in MEG research across three main aspects: disease treatment and practical applications, experimental foundations and technical understanding, and fundamental and advanced human cognition.

**Table 4. LLR algorithm of keywords**

| Cluster ID | Size | Silhouette | mean (Year) | Label (LLR) |
|---|---|---|---|---|
| 0 | 166 | 0.777 | 2000 | **magnetoencephalographic study (30872.12, 1.0E–4);** somatosensory evoked magnetic field (27484.47, 1.0E–4); secondary somatosensory cortex (12048.33, 1.0E–4); human somatosensory cortex (9521.74, 1.0E–4); somatosensory evoked field (9138.29, 1.0E–4) |
| 1 | 148 | 0.692 | 2008 | **working memory (18494.03, 1.0E–4);** visual working memory (9806.45, 1.0E–4); alpha oscillation (9272.64, 1.0E–4); gamma oscillation (8854.73, 1.0E–4); alpha activity (7946.06, 1.0E–4) |
| 2 | 140 | 0.72 | 2012 | **autism spectrum disorder (8895.29, 1.0E–4);** resting state (8223.7, 1.0E–4); brain network (8106.09, 1.0E–4); functional network (6978.89, 1.0E–4); subjective cognitive decline (6887.74, 1.0E–4) |
| 3 | 121 | 0.686 | 2006 | **meg source reconstruction (13786.33, 1.0E–4);** eeg source localization (11443.14, 1.0E–4); magnetoencephalographic study (11440.75, 1.0E–4); eeg data (10561.96, 1.0E–4); eeg source reconstruction (9387.68, 1.0E–4) |
| 4 | 116 | 0.753 | 2004 | **human auditory cortex (24568.45, 1.0E–4);** speech sound (11680.25, 1.0E–4); change detection (6788.74, 1.0E–4); mismatch negativity (6468.32, 1.0E–4); auditory evoked magnetic field (5146.55, 1.0E–4) |
| 5 | 106 | 0.805 | 2005 | **focal cortical dysplasia (11638.32, 1.0E–4);** presurgical evaluation (11508.79, 1.0E–4); epilepsy surgery (10280.01, 1.0E–4); intractable epilepsy (9642.18, 1.0E–4); temporal lobe epilepsy (9568.29, 1.0E–4) |
| 6 | 81 | 0.732 | 2007 | **left temporal lobe (3756.85, 1.0E–4);** complex word (2878.84, 1.0E–4); temporal area (2584.28, 1.0E–4); meg evidence (2074.99, 1.0E–4); facial expression (1917.5, 1.0E–4) |
| 7 | 43 | 0.826 | 2004 | **wada test (3383.81, 1.0E–4);** language dominance (3285.91, 1.0E–4); motor evoked-potential (2616.63, 1.0E–4); noninvasive alternative (2525.83, 1.0E–4); wada procedure (2359.21, 1.0E–4) |
| 8 | 4 | 0.985 | 1998 | **dipole source localization (74.5, 1.0E–4);** mean (74.5, 1.0E–4); theory (49.41, 1.0E–4); magnetoencephalography study (0.3, 1.0); magnetoencephalographic study (0.28, 1.0) |

First, disease treatment and practical application: cluster 2, which focuses on "autism spectrum disorder" (ASD), prominently emerges. ASD represents a heterogeneous group of neurodevelopmental disorders characterized by impairments in social communication. It stands as one of the most actively researched topics in psychiatry and neuroscience utilizing MEG technology. In recent years, research on the relationship between Autism Spectrum Disorder (ASD) and the human auditory cortex (cluster 4) and left temporal lobe (cluster 6) has been increasing. Some studies hypothesize that atypical low-level auditory perception in the cortex is the cause of language development abnormalities in children with ASD. For example, Arutiunian et al. (2023) used MEG to confirm the link and provided the first evidence for the association between the 40Hz auditory steady-state response in the left auditory cortex and language comprehension of children with ASD. As seen in cluster 5, focal cortical dysplasia (FCDs) belong to the broad spectrum of malformations of cortical development (MCDs) (Alexopoulos et al., 2006) and represent the most common structural brain lesion in children with drug-resistant focal epilepsies who undergo surgical treatment (Ikemoto et al., 2022). Guerrini et al. (2015) found that inadequate treatment options and difficulties in accessing and achieving successful surgical outcomes for patients with refractory epilepsy resulting from focal cortical dysplasia (FCD) present significant clinical challenges. Improvements in the precision of surgical resection, a critical factor for treatment success, have been achieved through targeted approaches that account for the physiological aspects of epileptogenesis in FCD and the localization of functional regions. Techniques such as EEG, fMRI, and MEG prove valuable in guiding the placement of electrodes and the surgical approach, while high-frequency oscillations play a crucial role in delineating the extent of the epileptogenic dysplasia. Parkinson's disease (PD) is the second most common neurodegenerative disease in the world, characterized by muscle rigidity, bradykinesia, and resting tremors (Jankovic, 2008). Research has provided important insights into how MEG reveals brain physiology changes in PD, especially how neural abnormalities impact the specificity of PD in subcortical and cortical motor networks, which has become an important focus in this field. For example, Wilson et al. (2016) found that PD-specific neural abnormalities were important in identifying specific potential for future clinical trials. Additionally, the combination of EEG and MEG source analysis has received widespread attention in presurgical evaluations for epilepsy. For example, Aydin et al. (2017) combined MEG and EEG source analysis information along with the latest MRI technology to demonstrate that the integration of MEG source analysis technology and biomagnetic signal analysis was highly effective in assessing non-lesioned neocortical epilepsy, leading to a significant improvement in the success rate of epilepsy surgery. In recent years, several studies have highlighted how global or nodal topological changes in the brain network may correlate with clinical assessments of cognitive functions or disease progression. For example, Pesoli et al. (2022) found that sleep deprivation has a profound impact on the connectivity of resting-state networks, such as the default mode network and the attention network. Researchers utilized

MEG and cognitive tasks to investigate the effects of 24 hours of sleep deprivation (SD) on brain topology and cognitive function. These findings demonstrate that 24 hours of sleep deprivation (SD) has the ability to modify brain connectivity and specifically impacts cognitive domains regulated by distinct brain networks. Moreover, MEG has been applied to the study of several neurodegenerative diseases, such as Parkinson 's disease (Lopez et al., 2023), amyotrophic lateral sclerosis (Romano et al., 2022), and mild cognitive impairment (Gómez et al., 2018), among others. Many neuroscientists use MEG to measure indicators of brain flexibility that can help predict disease severity, particularly in neurodegenerative diseases, paving the way for its use in personalized medicine (Ning et al., 2022). Second, the experimental basis and technical cognition. As shown in cluster 0 "magnetoencephalographic study", the interpretation of natural speech processing in the auditory cortex constitutes a major challenge in cognitive neuroscience. Luo and Poeppel (2007) uses MEG to study the mechanisms of speech dynamics analysis in the human auditory cortex, which lays a foundation for further neurological research. The research in Cluster 3 demonstrates that MEG source reconstruction has been the subject of numerous studies. For example, Bezsudnova et al. (2022) found that optimizing the sensing volume of OPM sensors is beneficial for MEG source reconstruction. Moreover, Vinding and Oostenveld (2022) proposed a solution to the dilemma regarding the use of anatomical MRIs in MEG source analysis. The proposed approach involves substituting individual anatomical MRIs with personalized warped templates. These templates should serve as an adequate basis for conducting MEG source analysis, possessing a suitable degree of geometric resemblance to the original participants' MRIs. Cluster 8 "dipole source localization" indicates that dipole source localization is a commonly utilized technique in neuroscience and brain imaging research to determine the source or location of neural activity within the brain. It utilizes electrophysiological data, such as EEG or MEG, to estimate the localization of neural activity within the brain. Typically, neural activity is represented by simplified models known as dipoles, which describe the electrical activity of neurons (Michel & He, 2019). By analyzing the spatiotemporal characteristics of the electrical signals, Dipole Source Localization aids researchers identifying active regions in the brain, determining which areas are involved in information processing during specific tasks or stimuli. This technique finds broad applications in studying cognitive functions (Attal et al., 2012), neurological disorders (Faes et al., 2022), and beyond. The results of Dipole Source Localization contribute to a better understanding of the spatial distribution and temporal characteristics of neural activity in the brain. As demonstrated in cluster 7, the Wada test - also known as the intracarotid amobarbital procedure - is a standard practice in many medical centers worldwide for assessing language and memory functions in each hemisphere (Baxendale, 2009; Haag et al., 2008). Over the past decade, there has been a significant increase in research efforts focused on replacing the Wada test with noninvasive neuroimaging methods, such as fMRI, MEG, and alternatives like functional transcranial Doppler sonography (Bauer et al., 2014; Papanicolaou et al., 2014). Kemp et al. (2018) investigated the concordance between Wada test and two imaging techniques, MEG and fMRI, in studies preceding epileptic surgery.

Third, the human basis and advanced cognition. As demonstrated in cluster 1 "working memory", working memory (WM) plays a foundational role in various cognitive abilities, encompassing goal-directed behaviors, inference, calculations, and decision-making (D'Esposito & Postle, 2015). While the neural mechanisms underlying it are still debated (Constantinidis et al., 2018), an expanding body of research consistently demonstrated a strong association between WM and oscillatory brain signals (Kornblith et al., 2016). Alpha oscillations play a role in working memory. For instance, Noguchi and Kakigi (2020) found visual working memory (vWM) is a critical cognitive ability essential for various tasks, although its neural basis remains elusive. While previous research has predominantly examined the θ (4–7 Hz) and γ (30–60 Hz) frequency bands as potential substrates of vWM, our study shows that temporal signals within the α (8–12 Hz) and β (13–30 Hz) bands can accurately predict vWM capacity. Using MEG, we recorded neural activity from healthy human participants performing a classic vWM task (change detection). Through analysis of peak interval (IPI) variations in oscillatory signals and the modulation of WM load (number of items to remember, 1-6), our findings demonstrate reduced load dependence of IPIs in parietal and frontal regions, indicating accelerated α/β rhythms during storage of multiple items in vWM. Importantly, this reduction in IPI correlates positively with individual vWM capacity, especially in the frontal cortex. Collectively, our results suggest that vWM is characterized by changes in oscillation frequencies within the human cerebral cortex. MEG has been widely employed to study the spatiotemporal dynamics underlying higher cognitive processes and age-related differences, particularly in research on working memory. For instance, Proskovec et al. (2016) found that both young and older individuals exhibited widespread neural activity in the left temporal lobe during encoding and maintenance of working memory, only older individuals engaged the right frontal cortex in WM processing. Functional connectivity and reduced mutual coupling are closely associated with MEG. MEG is an excellent tool for studying functional connectivity in large-scale brain activity by detecting interactions between different brain regions and revealing the processes of information processing in the brain. The high temporal and spatial resolution of MEG provides comprehensive data covering the entire brain, describing dynamic changes between neuronal populations and playing a crucial role in studying correlations and dependencies among different brain regions (Bastos & Schoffelen, 2016). However, functional connectivity also has some limitations; for instance, MEG sensors record activation in each brain region, yet the magnitude is determined by the direction and location of activated neuron groups. Therefore, activation in a single brain area can potentially create spurious connections across all MEG sensors, leading to delays and errors in source analysis (Palva et al., 2018). Mutual coupling primarily refers to the theoretical mechanism by which neural oscillations generate interactions between brain regions, and MEG is an ideal tool for non-invasively detecting this phenomenon (Marzetti et al., 2019). For example, it is noteworthy that Marzetti et al. (2019), distinguished between functional connectivity and power to demonstrate genuine phase coupling between brain regions. The connectivity theory of MRI function and anatomy supported MEG connectivity findings in both resting-state and visual spatial templates. Therefore, in recent years, research on functional connectivity and mutual coupling in MEG has made significant strides, allowing for the study of advanced cognitive functions and changes in brain rhythm with the aid of more integrated modal tools.

Although MEG is an advanced neurophysiological imaging tool and has become a research instrument for exploring the neural dynamics underlying the brain and psychiatric disorders, it still faces limitations such as complexity and inaccuracy during application. Currently, MEG has the following limitations:

First, most existing MEG studies are conducted in laboratory environments with significant spatial constraints and limited interaction contexts. Even the latest wearable MEG technology can only detect electromagnetic signals from the human brain during moderate activities, and the data obtained under such conditions lack necessary ecological validity.

Second, MEG is a completely passive technology. Unlike eye-tracking techniques that depend on active eye movements, passive research can reduce individual engagement, potentially leading to fatigue and boredom. These factors can influence fluctuations in brain magnetic signals and affect experimental results.

Third, during source analysis in MEG, functional connections can generate "ghost" interactions, creating false connections that modern technology has not fully overcome, necessitating ongoing research.

Based on the aforementioned limitations and the outcomes of the clustering analysis, this study outlines the following prospects for MEG research:

First, the volume of MEG studies has grown rapidly over the past decade, and although research fervor has slightly tapered off in the past year, overall, this upward trajectory is anticipated to persist into the next decade. With ongoing attention from clinical cognitive neuroscience on brain dynamics, the utilization of MEG is expected to increase. In line with the principles and limitations of MEG, researchers are expected to develop new, more convenient, and less spatially restricted methods for its use. As its applications advance, research costs are also anticipated to decrease.

Second, further utilization of MEG for studying functional connections and network activity in the human brain holds promise for advancing cognitive neuroscience in psychology and clinical medicine. Future research should further explore the applied role of MEG in diseases such as epilepsy, autism spectrum disorders, Alzheimer's disease, Parkinson's disease, ADHD, and other diseases where symptoms stem from abnormalities across multiple brain regions. With its expanding applications, MEG may emerge in new fields in the coming years and continue to gain attention from the scientific community and researchers.

Third, enhancing the integration of MEG with other instruments diversifies research methodologies and fosters multidisciplinary collaborations. Most current research integrates MEG with other technologies such as EEG or fMRI, with MEG compensating for the lower spatial resolution of EEG. Diversifying MEG research methods primarily involves leveraging artificial intelligence technology to expand the diversity of measurement data and detection methods in MEG. For example, leveraging artificial intelligence can automate MEG measurements, akin to autonomous driving technology in cars. Automated measurements enhance data accuracy by minimizing the influence of extraneous variables such as the Hawthorne effect.

## 5. Conclusions

Through the first large-scale bibliometric analysis of MEG over the past 28 years, the following conclusions can be drawn:

(1) From the statistics analysis of literature distribution characteristics, it was found that 8,331 articles authored by 18,592 researchers from 4,072 institutions are published in 839 journals across 92 countries or regions. Moreover, the temporal distribution of the literature also mirrors the overall global trend in MEG research. There were a total of 8,331 English articles and review articles on MEG published worldwide between 1995 and 2022. The annual output of relevant publications has expanded dramatically, drawing increasing attention from scholars to MEG. In terms of research output, the United States unrivaled. Aalto University holds the highest number of publications in the field of MEG, underscoring its outstanding academic standing. Ryusuke Kakigi, from Aalto University, is the author with the most papers, highlighting significant achievements in the field of MEG. Neuroimage is the journal with the highest number of articles, establishing MEG as a distinctive focus within the journal.

(2) From the analysis of keyword hotspots, it was found that recent keywords with significant citation burst strength include "resting state" (14.66, 2014–2022), "network" (33.23, 2017–2022), "functional connectivity" (39.71, 2017–2022), "phase" (15.46, 2017–2022), "connectivity" (29.76, 2019–2022), and "oscillation" (19.62, 2019–2022). Disease treatment and practical application, experimental foundations and technical insights, and fundamental and advanced human cognition will be three specific directions for future research.

**Data availability** The data are available from the corresponding author upon reasonable request.

## Declarations

**Competing interests** The authors declare that the research was conducted in the absence of any commercial or financial relationships that could be construed as a potential conflict of interest.

## References

Alexopoulos, A. V., Kotagal, P., Loddenkemper, T., Hammel, J., & Bingaman, W. E. (2006). Long-term results with vagus nerve stimulation in children with pharmacoresistant epilepsy. *Seizure, 15*(7), 491–503.

Arutiunian, V., Arcara, G., Buyanova, I., Davydova, E., Pereverzeva, D., …, Dragoy, O. (2023). Neuromagnetic 40 Hz auditory steady-state response in the left auditory cortex is related to language comprehension in children with autism spectrum disorder. *Progress in Neuro-psychopharmacology & Biological Psychiatry, 122*, 110690.

Attal, Y., Maess, B., Friederici, A., & David, O. (2012). Head models and dynamic causal modeling of subcortical activity using magnetoencephalographic/electroencephalographic data. *Reviews in the Neurosciences, 23*(1), 85–95.

Aydin, Ü., Rampp, S., Wollbrink, A., Kugel, H., Cho, J. H., Knösche, T. R., Grova, C., Wellmer, J., & Wolters, C. H. (2017). Zoomed MRI guided by combined EEG/MEG source analysis: A multimodal approach for optimizing presurgical epilepsy work-up and its application in a multi-focal epilepsy patient case study. *Brain Topography, 30*(4), 417–433.

Baillet, S. (2017). Magnetoencephalography for brain electrophysiology and imaging. *Nature*

*Neuroscience, 20*(3), 327–339.

Bastos, A. M., & Schoffelen, J. M. (2016). A tutorial review of functional connectivity analysis methods and their interpretational pitfalls. *Frontiers in Systems Neuroscience, 9*, 175.

Bauer, P. R., Reitsma, J. B., Houweling, B. M., Ferrier, C. H., & Ramsey, N. F. (2014). Can fMRI safely replace the Wada test for preoperative assessment of language lateralisation? A meta-analysis and systematic review. *Journal of Neurology, Neurosurgery and Psychiatry, 85*(5), 581–588.

Baxendale, S. (2009). The Wada test. *Current Opinion in Neurology, 22*(2), 185–189.

Bezsudnova, Y., Koponen, L. M., Barontini, G., Jensen, O., & Kowalczyk, A. U. (2022). Optimising the sensing volume of OPM sensors for MEG source reconstruction. *NeuroImage, 264*, 119747.

Boto, E., Holmes, N., Leggett, J., Roberts, G., Shah, V., …, Brookes, M. J. (2018). Moving magnetoencephalography towards real-world applications with a wearable system. *Nature, 555,* 657–661.

Brookes, M. J., Leggett, J., Rea, M., Hill, R. M., Holmes, N., Boto, E., & Bowtell, R. (2022). Magnetoencephalography with optically pumped magnetometers (OPM-MEG): The next generation of functional neuroimaging. *Trends in Neurosciences, 45*(8), 621–634.

Bullmore, E., & Sporns, O. (2009). Complex brain networks: Graph theoretical analysis of structural and functional systems. *Nature Reviews Neuroscience, 10*(3), 186–198.

Chen, C. M. (2006). CiteSpace II: Detecting and visualizing emerging trends and transient patterns in scientific literature. *Journal of the American for Information Science and Technology, 57*(3), 359–377.

Cohen, D. (1972). Magnetoencephalography: Detection of the brain's electrical activity with a superconducting magnetometer. *Sciences, 175*(4022), 664–666.

Constantinidis, C., Funahashi, S., Lee, D., Murray, J. D., Qi, X. L., Wang, M., & Arnsten, A. F. (2018). Persistent spiking activity underlies working memory. *Journal of Neuroscience, 38*(32), 7020–7028.

D'Esposito, M., & Postle, B. R. (2015). The cognitive neuroscience of working memory. *Annual Review of Psychology, 66*, 115–142.

Ding, X., & Yang, Z. (2022). Knowledge mapping of platform research: A visual analysis using VOSviewer and CiteSpace. *Electronic Commerce Research*, *22*, 787–809.

Faes, A., Vantieghem, I., & Van Hulle, M. M. (2022). Neural networks for directed connectivity estimation in source-reconstructed EEG data. *Applied Sciences, 12*(6), 2889.

Gómez, C., Juan-Cruz, C., Poza, J., Ruiz-Gómez, S. J., Gomez-Pilar, J., …, Hornero, R. (2018). Alterations of effective connectivity patterns in mild cognitive impairment: An MEG study. *Journal of Alzheimer's Disease, 65*(3), 843–854.

Guerrini, R., Duchowny, M., Jayakar, P., Krsek, P., Kahane, P., …, Blumcke, I. (2015). Diagnostic methods and treatment options for focal cortical dysplasia. *Epilepsia, 56*(11), 1669–1686.

Haag, A., Knake, S., Hamer, H. M., Boesebeck, F., Freitag, H., …, Rosenow, F. (2008). The Wada test in Austrian, Dutch, German, and Swiss epilepsy centers from 2000 to 2005: A review of 1421 procedures. *Epilepsy & Behavior, 13*(1), 83–89.

Hauk, O. (2020). Human cognitive neuroscience as it is taught. *Frontiers in Psychology, 11*, 587922.

Ikemoto, S., von Ellenrieder, N., & Gotman, J. (2022). Electroencephalography–functional magnetic resonance imaging of epileptiform discharges: Noninvasive investigation of the whole brain. *Epilepsia*, *63*(11), 2725–2744.

Jankovic, J. (2008). Parkinson's disease: Clinical features and diagnosis. *Journal of Neurology, Neurosurgery and Psychiatry, 79*(4), 368–376.

Johnson, C. N., Schwindt, P. D. D., & Weisend, M. (2013). Multi-sensor magnetoencephalography with atomic magnetometers. *Physics in Medicine & Biology, 58*(17), 6065–6077.

Kemp, S., Prendergast, G., Karapanagiotidis, T., Baker, G., Kelly, T. P., …, & Keller, S. S. (2018). Concordance between the Wada test and neuroimaging lateralization: Influence of imaging modality (fMRI and MEG) and patient experience. *Epilepsy & Behavior, 78*, 155–160.

Kharkar, S., & Knowlton, R. (2015). Magnetoencephalography in the presurgical evaluation of epilepsy. *Epilepsy & Behavior, 46*, 19–26.

Kornblith, S., Buschman, T. J., & Miller, E. K. (2016). Stimulus load and oscillatory activity in higher cortex. *Cerebral Cortex, 26*(9), 3772–3784.

Kovačić, N., & Petrak, J. (2022). Three decades of the *Croatian Medical Journal* - Can small non-profit journal compete in the bibliometrics arena? *Croatian Medical Journal, 63*(6), 501–507.

Lopez, E. T., Minino, R., Liparoti, M., Polverino, A., Romano, A., …, Sorrentino, P. (2023). Fading of brain network fingerprint in Parkinson's disease predicts motor clinical impairment. *Human Brain Mapping, 44*(3), 1239–1250.

Luo, H., & Poeppel, D. (2007). Phase patterns of neuronal responses reliably discriminate speech in human auditory cortex. *Neuron, 54*(6), 1001–1010.

Marzetti, L., Basti, A., Chella, F., D'Andrea, A., Syrjälä, J., & Pizzella, V. (2019). Brain functional connectivity through phase coupling of neuronal oscillation: A perspective from magnetoencephalography. *Frontiers in Neuroscience, 13*, 964.

Michel, C. M., & He, B. (2019). EEG source localization. *Handbook of Clinical Neurology, 160*, 85–101.

Ning, Y. Z., Zheng, S. S., Feng, S. T., Li, K. S., & Jia, H. X. (2022). Altered functional connectivity and topological organization of brain networks correlate to cognitive impairments after sleep deprivation.

*Nature and Science of Sleep, 14*, 1285–1297.

Noguchi, Y., & Kakigi, R. (2020). Temporal codes of visual working memory in the human cerebral cortex: Brain rhythms associated with high memory capacity. *NeuroImage, 222*, 117294.

Palva, J. M., Wang, S. H., Palva, S., Zhigalov, A., Monto, S., Brookes, M. J., …, Jerbi, K. (2018). Ghost interactions in MEG/EEG source space: A note of caution on inter-areal coupling measures. *NeuroImage, 173*, 632–643.

Pan, H. T., Xi, Z. Q., Yu, X. T., Sun, X. Q., Wei, X. Q., & Wang, K. (2021). Knowledge mapping analysis of international research on acupuncture for low back pain using bibliometrics. *Journal of Pain Research, 14*, 3733–3746.

Papanicolaou, A. C., Rezaie, R., Narayana, S., Choudhri, A. F., Wheless, J. W., …, Boop, F. A. (2014). Is it time to replace the Wada test and put awake craniotomy to sleep? *Epilepsia, 55*(5), 629–632.

Pesoli, M., Rucco, R., Liparoti, M., Lardone, A., D'Aurizio, G., …, Sorrentino, P. (2022). A night of sleep deprivation alters brain connectivity and affects specific executive functions. *Neurological Sciences, 43*(2), 1025–1034.

Proskovec, A. L., Heinrichs-Graham, E., & Wilson, T. W. (2016). Aging modulates the oscillatory dynamics underlying successful working memory encoding and maintenance. *Human Brain Mapping, 37*(6), 2348–2361.

Romano, A., Lopez, E. T., Liparoti, M., Polverino, A., Minino, R., …, Sorrentino, P. (2022). The progressive loss of brain network fingerprints in Amyotrophic Lateral Sclerosis predicts clinical impairment. *NeuroImage: Clinical, 35*, 103095.

Van Eck, N., & Waltman, L. (2010). Software survey: VOSviewer, a computer program for bibliometric mapping. *Scientometrics, 84*(2), 523–538.

Vinding, M. C., & Oostenveld, R. (2022). Sharing individualised template MRI data for MEG source reconstruction: A solution for open data while keeping subject confidentiality. *Neuroimage, 254*, 119165.

Wilson, T. W., Heinrichs-Graham, E., Proskovec, A. L., & McDermott, T. J. (2016). Neuroimaging with magnetoencephalography: A dynamic view of brain pathophysiology. *Translational Research, 175*, 17–36.

Yan, W. W., Zheng, K. Y., Weng, L. M., Chen, C. C., Kiartivich, S., Jiang, X., Su, X., Wang, Y. L., & Wang, X. Q. (2020). Bibliometric evaluation of 2000–2019 publications on functional near-infrared spectroscopy. *NeuroImage, 220*, 117121.

**Supplementary Material**

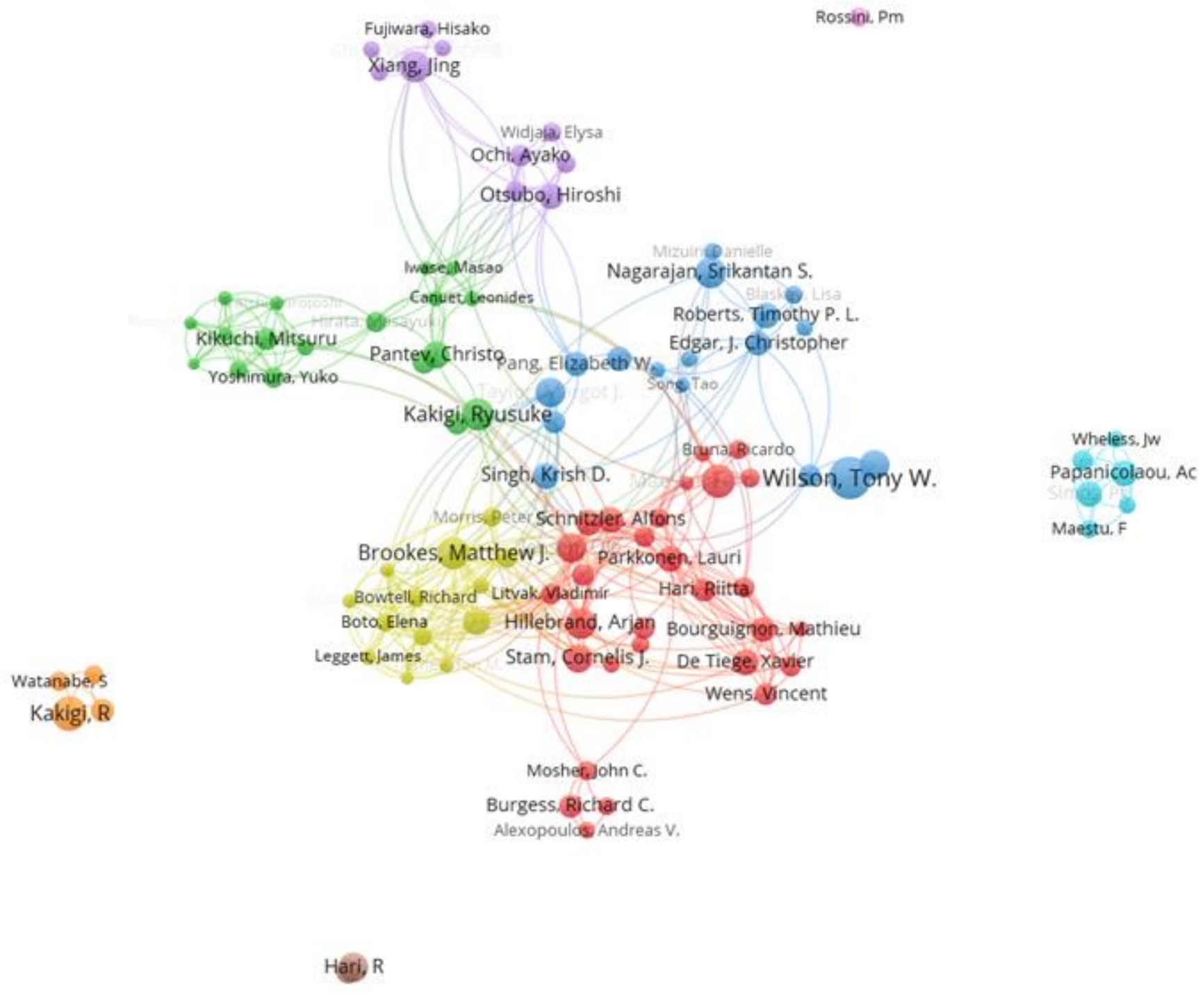

**Supplementary Figure 1. Cluster view of authors of MEG research**

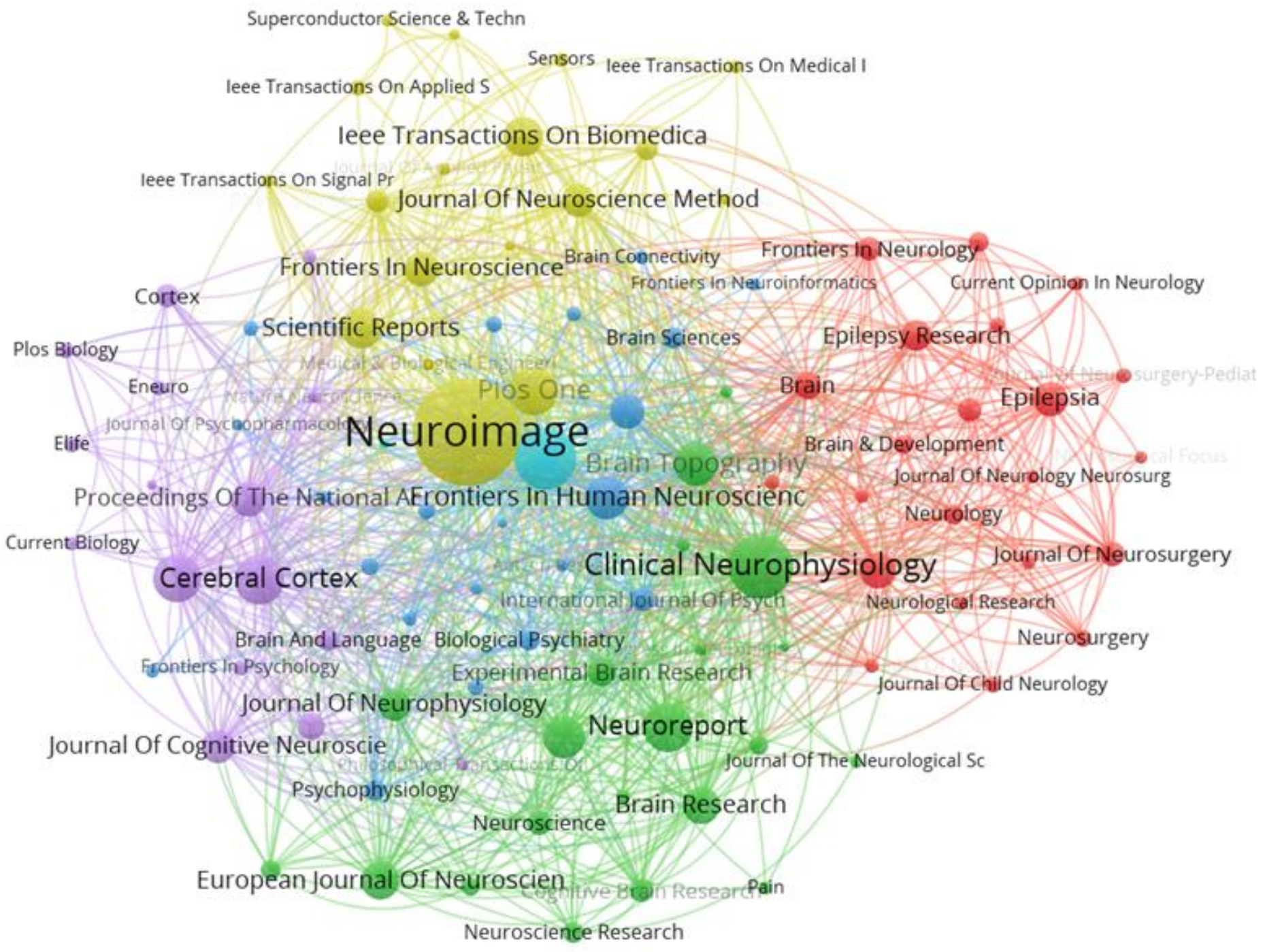

**Supplementary Figure 2. Cluster view of journals published in MEG research**

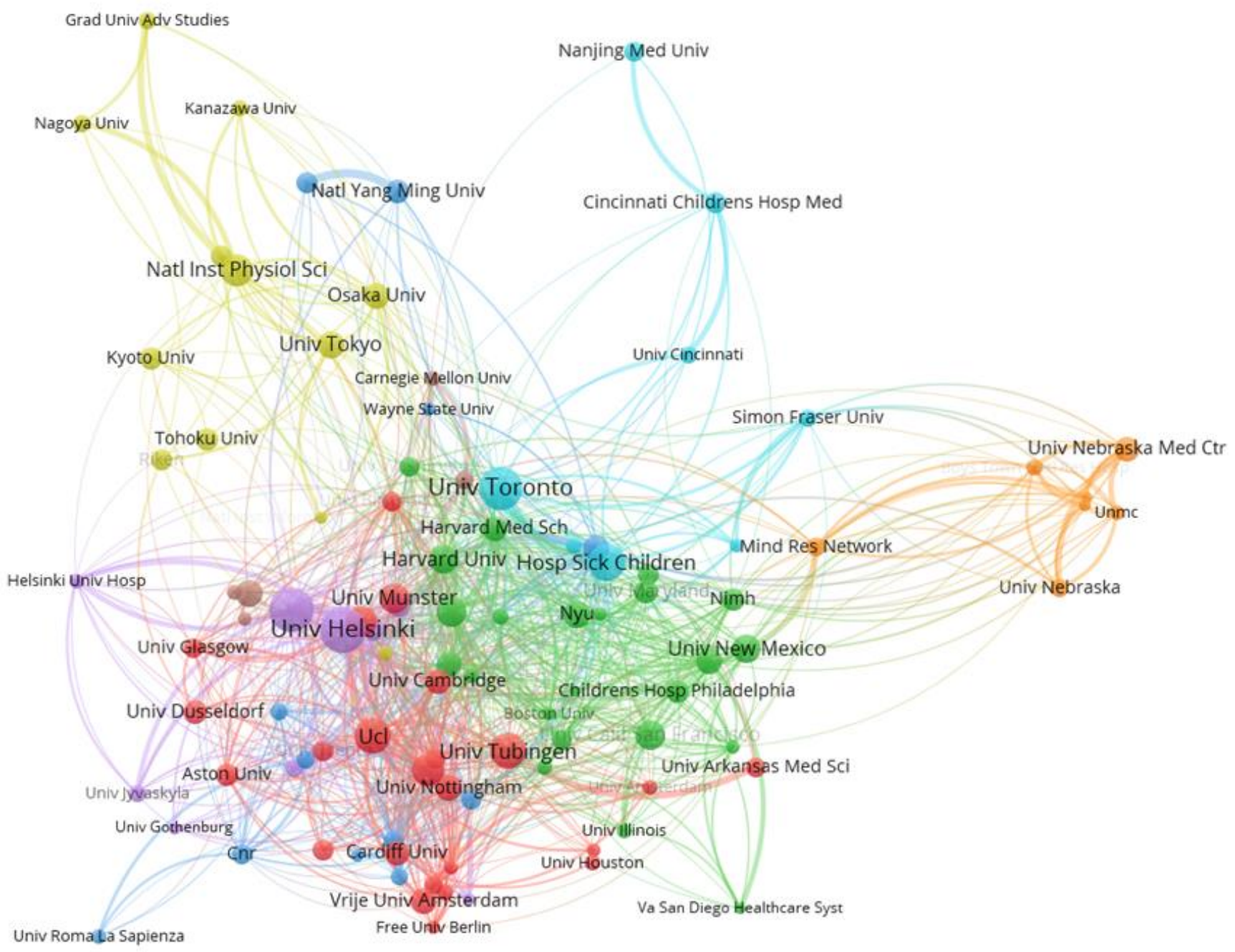

**Supplementary Figure 3. Cluster view of author institutions of MEG research**

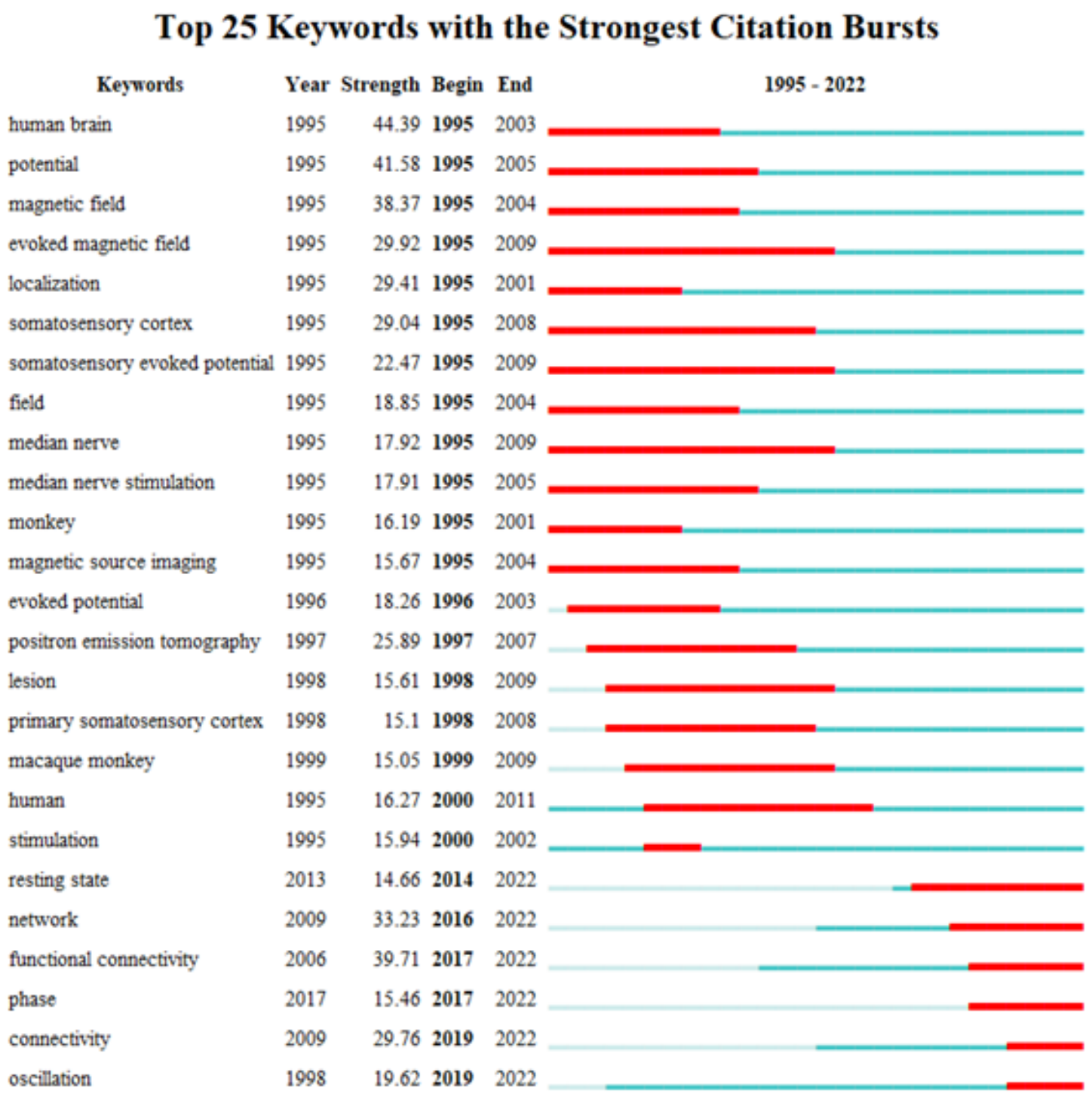

**Supplementary Figure 4. Burst view of keywords**

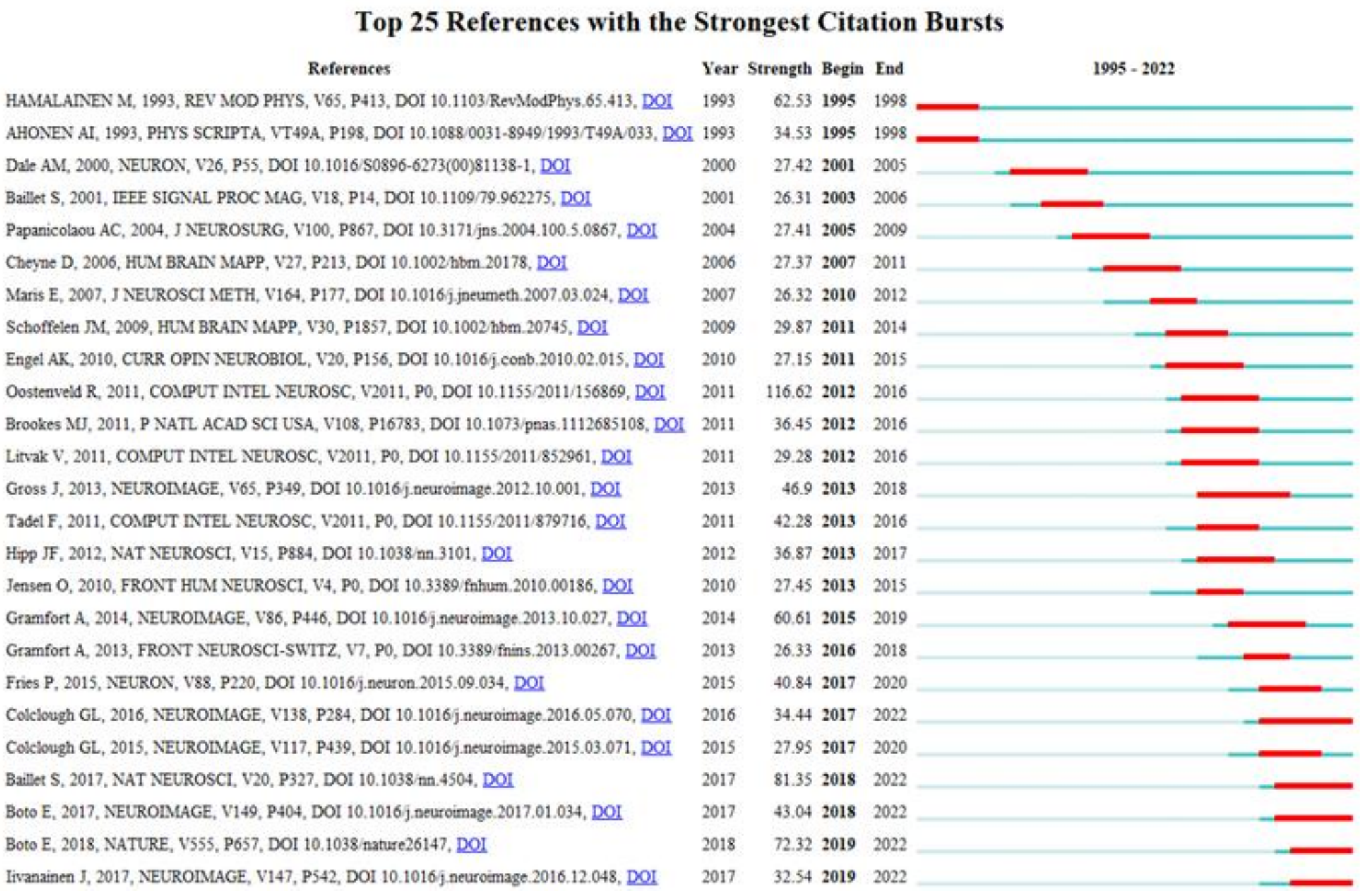

**Top 25 References with the Strongest Citation Bursts**

| References | Year | Strength | Begin | End | 1995 - 2022 |
|---|---|---|---|---|---|
| HAMALAINEN M, 1993, REV MOD PHYS, V65, P413, DOI 10.1103/RevModPhys.65.413, DOI | 1993 | 62.53 | **1995** | 1998 | |
| AHONEN AI, 1993, PHYS SCRIPTA, VT49A, P198, DOI 10.1088/0031-8949/1993/T49A/033, DOI | 1993 | 34.53 | **1995** | 1998 | |
| Dale AM, 2000, NEURON, V26, P55, DOI 10.1016/S0896-6273(00)81138-1, DOI | 2000 | 27.42 | **2001** | 2005 | |
| Baillet S, 2001, IEEE SIGNAL PROC MAG, V18, P14, DOI 10.1109/79.962275, DOI | 2001 | 26.31 | **2003** | 2006 | |
| Papanicolaou AC, 2004, J NEUROSURG, V100, P867, DOI 10.3171/jns.2004.100.5.0867, DOI | 2004 | 27.41 | **2005** | 2009 | |
| Cheyne D, 2006, HUM BRAIN MAPP, V27, P213, DOI 10.1002/hbm.20178, DOI | 2006 | 27.37 | **2007** | 2011 | |
| Maris E, 2007, J NEUROSCI METH, V164, P177, DOI 10.1016/j.jneumeth.2007.03.024, DOI | 2007 | 26.32 | **2010** | 2012 | |
| Schoffelen JM, 2009, HUM BRAIN MAPP, V30, P1857, DOI 10.1002/hbm.20745, DOI | 2009 | 29.87 | **2011** | 2014 | |
| Engel AK, 2010, CURR OPIN NEUROBIOL, V20, P156, DOI 10.1016/j.conb.2010.02.015, DOI | 2010 | 27.15 | **2011** | 2015 | |
| Oostenveld R, 2011, COMPUT INTEL NEUROSC, V2011, P0, DOI 10.1155/2011/156869, DOI | 2011 | 116.62 | **2012** | 2016 | |
| Brookes MJ, 2011, P NATL ACAD SCI USA, V108, P16783, DOI 10.1073/pnas.1112685108, DOI | 2011 | 36.45 | **2012** | 2016 | |
| Litvak V, 2011, COMPUT INTEL NEUROSC, V2011, P0, DOI 10.1155/2011/852961, DOI | 2011 | 29.28 | **2012** | 2016 | |
| Gross J, 2013, NEUROIMAGE, V65, P349, DOI 10.1016/j.neuroimage.2012.10.001, DOI | 2013 | 46.9 | **2013** | 2018 | |
| Tadel F, 2011, COMPUT INTEL NEUROSC, V2011, P0, DOI 10.1155/2011/879716, DOI | 2011 | 42.28 | **2013** | 2016 | |
| Hipp JF, 2012, NAT NEUROSCI, V15, P884, DOI 10.1038/nn.3101, DOI | 2012 | 36.87 | **2013** | 2017 | |
| Jensen O, 2010, FRONT HUM NEUROSCI, V4, P0, DOI 10.3389/fnhum.2010.00186, DOI | 2010 | 27.45 | **2013** | 2015 | |
| Gramfort A, 2014, NEUROIMAGE, V86, P446, DOI 10.1016/j.neuroimage.2013.10.027, DOI | 2014 | 60.61 | **2015** | 2019 | |
| Gramfort A, 2013, FRONT NEUROSCI-SWITZ, V7, P0, DOI 10.3389/fnins.2013.00267, DOI | 2013 | 26.33 | **2016** | 2018 | |
| Fries P, 2015, NEURON, V88, P220, DOI 10.1016/j.neuron.2015.09.034, DOI | 2015 | 40.84 | **2017** | 2020 | |
| Colclough GL, 2016, NEUROIMAGE, V138, P284, DOI 10.1016/j.neuroimage.2016.05.070, DOI | 2016 | 34.44 | **2017** | 2022 | |
| Colclough GL, 2015, NEUROIMAGE, V117, P439, DOI 10.1016/j.neuroimage.2015.03.071, DOI | 2015 | 27.95 | **2017** | 2020 | |
| Baillet S, 2017, NAT NEUROSCI, V20, P327, DOI 10.1038/nn.4504, DOI | 2017 | 81.35 | **2018** | 2022 | |
| Boto E, 2017, NEUROIMAGE, V149, P404, DOI 10.1016/j.neuroimage.2017.01.034, DOI | 2017 | 43.04 | **2018** | 2022 | |
| Boto E, 2018, NATURE, V555, P657, DOI 10.1038/nature26147, DOI | 2018 | 72.32 | **2019** | 2022 | |
| Iivanainen J, 2017, NEUROIMAGE, V147, P542, DOI 10.1016/j.neuroimage.2016.12.048, DOI | 2017 | 32.54 | **2019** | 2022 | |

**Supplementary Figure 5. Burst view of references**

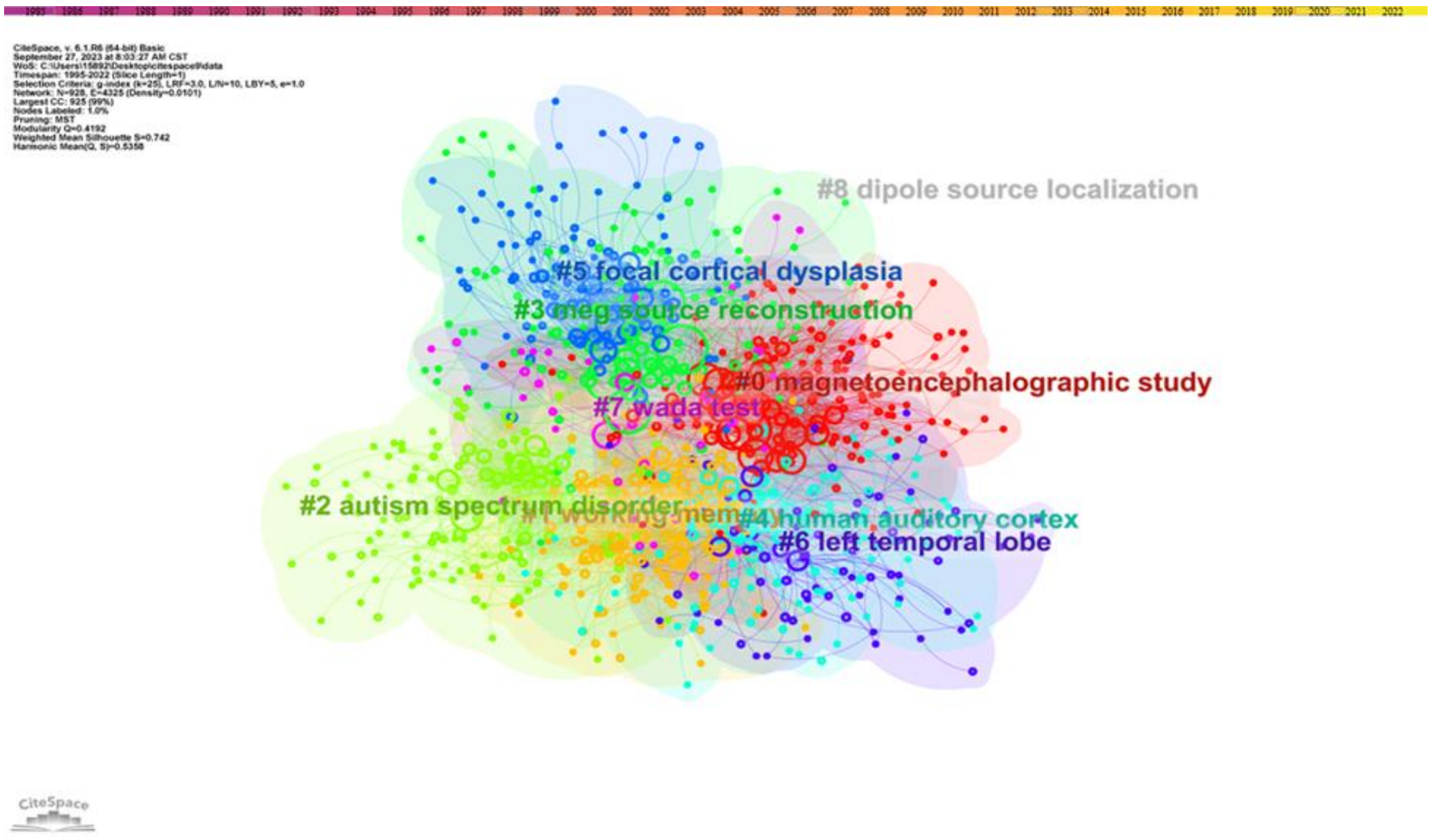

**Supplementary Figure 6. Keywords in MEG study cluster view via CiteSpace**

**Supplementary Table 1 Top 10 countries in terms of publication of MEG research**

| Rank | Country | Publication (%) |
|---|---|---|
| 1 | United States of American | 2796 (33.56) |
| 2 | Germany | 1573 (18.88) |
| 3 | Japan | 1300 (15.60) |
| 4 | United Kingdom | 1235 (14.82) |
| 5 | Canada | 810 (9.72) |
| 6 | Finland | 787 (9.45) |
| 7 | People's Republic of China | 590 (7.08) |
| 8 | Netherlands | 511 (6.13) |
| 9 | France | 456 (5.47) |
| 10 | Italy | 446 (5.35) |

**Supplementary 2 Top 10 institutions in the number of publications on MEG research**

| Rank | Institution | Publication (%) |
|---|---|---|
| 1 | Aalto University | 538 (6.46) |
| 2 | University of Toronto | 465 (5.58) |
| 3 | University of Helsinki | 418 (5.02) |
| 4 | University of California System | 399 (4.79) |
| 5 | University of London | 392 (4.71) |
| 6 | Harvard University | 369 (4.43) |
| 7 | Udice French Research Universities | 353 (4.24) |
| 8 | Helsinki University Central Hospital | 338 (4.06) |
| 9 | University College London | 327 (3.93) |
| 10 | Hospital for Sick Children (Sick Kids) | 307 (3.69) |

**Supplementary Table 3. Top 10 references cited**

| Rank | Reference | Citation (WOS) |
|---|---|---|
| 1 | Bullmore, E., & Sporns, O. (2009). Complex brain networks: Graph theoretical analysis of structural and functional systems. *Nature Reviews. Neuroscience*, *10*(3), 186–198. | 7105 |
| 2 | Tadel, F., Baillet, S., Mosher, J. C., Pantazis, D., & Leahy, R. M. (2011). Brainstorm: A user-friendly application for MEG/EEG analysis. *Computational Intelligence and Neuroscience*, *2011*, 879716. | 1943 |
| 3 | Näätänen, R., Paavilainen, P., Rinne, T., & Alho, K. (2007). The mismatch negativity (MMN) in basic research of central auditory processing: a review. *Clinical Neurophysiology: Official Journal of the International Federation of Clinical Neurophysiology*, *118*(12), 2544–2590. | 1751 |
| 4 | Bassett, D. S., & Bullmore, E. (2006). Small-world brain networks. *the Neuroscientist: A Review Journal Bringing Neurobiology, Neurology and Psychiatry*, *12*(6), 512–523. | 1682 |
| 5 | Jensen, O & Mazaheri, A. (2010). Shaping functional architecture by oscillatory alpha activity: Gating by inhibition. *frontiers in Human Neuroscience*, *4*, 186 | 1663 |
| 6 | Rodriguez, E., George, N., Lachaux, J. P., Martinerie, J., | 1465 |
| | Renault, B., & Varela, F. J. (1999). Perception's shadow: Long-distance synchronization of human brain activity. *Nature*, *397*(6718), 430–433. | |
| 7 | Kao, M. H., Mandal, A., Lazar, N., & Stufken, J. (2009). Multi-objective optimal experimental designs for event-related fMRI studies. *Neuroimage*, *44*(3), 849–856. | 1356 |
| 8 | Van Essen, D. C., Ugurbil, K., Auerbach, E., Barch, D., Behrens, T. E., Bucholz, R., …&WU-Minn HCP Consortium (2012). The Human Connectome Project: a data acquisition perspective. *Neuroimage*, *62*(4), 2222–2231. | 1270 |
| 9 | Gross, J., Kujala, J., & Hamalainen, M. (2001). Dynamic imaging of coherent sources: sStudying neural interactions in the human brain. *Proceedings of the National Academy of Sciences*, *98*(2), 694–699. | 1262 |
| 10 | Dale, A. M., Liu, A. K., Fischl, B. R., Buckner, R. L., Belliveau, J. W., Lewine, J. D., & Halgren, E. (2000). Dynamic statistical parametric mapping: Combining fMRI and MEG for high-resolution imaging of cortical activity. *Neuron*, *26*(1), 55–67. | 1185 |